\documentclass[12pt]{article}
\usepackage{epsf,amsmath,amssymb,graphicx,scalefnt}
\usepackage{cite}
\usepackage{rotating}

\allowdisplaybreaks

\newcommand{\gammas}{\Gamma^{\raisebox{0.2mm}{$\scriptstyle s$}}}

\title{
  \vskip-3cm{\baselineskip14pt
    \begin{flushleft}
      \normalsize SFB/CPP-13-83 \\
      \normalsize TTP13-034 \\
    \end{flushleft}}
  \vskip1.5cm
  Three-loop QCD corrections to $B_s \to \mu^+ \mu^-$ }
\author{Thomas Hermann$^{(a)}$, Miko{\l}aj Misiak$^{(b, c)}$ and
  Matthias Steinhauser$^{(a)}$\\[2em]
  {\it (a) Institut f\"ur Theoretische Teilchenphysik}\\
  {\it Karlsruhe Institute of Technology (KIT), D-76128 Karlsruhe, Germany}\\[1em]
  {\it (b) Institute of Theoretical Physics, University of Warsaw,}\\
  {\it Ho\.za 69, PL-00-681 Warsaw, Poland}\\[1em]
  {\it (c) Theory Division, CERN, CH-1211 Geneva 23, Switzerland}\\[1em]
  }

\date{}

\begin{document}
\maketitle

%- {{{ abstract:

\begin{abstract}
  The decay $B_s \to \mu^+ \mu^-$ in the Standard Model is generated by
      the well-known $W$-box and $Z$-penguin diagrams that give rise to an
      effective quark-lepton operator $Q_A$ at low energies. We compute QCD
      corrections of order $\alpha_s^2$ to its Wilson coefficient $C_A$.  It
      requires performing three-loop matching between the full and effective
      theories. Including the new corrections makes $C_A$ more stable with
      respect to the matching scale $\mu_0$ at which the top-quark mass and
      $\alpha_s$ are renormalized. The corresponding uncertainty in
      $|C_A|^2$ gets reduced from around $1.8\%$ to less than $0.2\%$.  
      Our results are directly applicable to all the $B_{s(d)} \to
      \ell^+ \ell^-$ decay modes.
\end{abstract}

\thispagestyle{empty}

%- }}}

\newpage

%- {{{ Introduction:

\section{Introduction}

The decay $B_s \to \mu^+ \mu^-$ is well known as a probe of
  physics beyond the Standard Model (SM). Recently, it has attracted
a lot of attention since the LHCb and the CMS experiments at the CERN LHC have
provided first measurements of its branching 
ratio~\cite{Aaij:2012nna,Aaij:2013aka,Chatrchyan:2013bka}. Their 
current results for the average time-integrated branching
  ratio read
\begin{eqnarray}
  \overline{{\mathcal B}}(B_s \to \mu^+ \mu^-) 
  = \left(\, 2.9^{\, +1.1}_{\,-1.0}\, \right) \times 10^{-9}\,,
  && \mbox{LHCb~\cite{Aaij:2013aka}},
  \nonumber \\[1mm]
  \overline{{\mathcal B}}(B_s \to \mu^+ \mu^-) 
  = \left(\, 3.0^{\, +1.0}_{\,-0.9}\, \right) \times 10^{-9}\,,
  && \mbox{CMS~\cite{Chatrchyan:2013bka}},
  \label{eq::BRexp}
\end{eqnarray}
which leads to the weighted average~\cite{CMS.LHCb.average:2013xxx}
\begin{equation}
  \overline{{\mathcal B}}(B_s \to \mu^+ \mu^-) 
  = \left(\, 2.9 \pm 0.7 \right) \times 10^{-9}\,.
\end{equation}
Previous upper limits can be found in 
Refs.~\cite{Abazov:2010fs,Aaltonen:2011fi,Chatrchyan:2012rga,Aaij:2012ac,Aad:2012pn}.
Although the experimental uncertainties are still quite large, they are expected
to get significantly reduced within the next few years.

As far as the theory side is concerned, the $B_s$ meson decay into two muons
is quite clean. In fact, the only relevant quantity that needs to be calculated at the
leading order in $\alpha_{em}$ and cannot be determined within perturbation
theory is the leptonic decay constant $f_{B_s}$. Its square enters the
branching ratio as a multiplicative factor. Recent progress in the
determination of $f_{B_s}$ from lattice 
calculations~\cite{Dimopoulos:2011gx,McNeile:2011ng, Bazavov:2011aa,
Bernardoni:2012fd, Carrasco:2012ps, Dowdall:2013tga} gives a motivation for
improving the perturbative ingredients, in particular the two-loop
electroweak~\cite{Bobeth:2013tba} and the three-loop QCD corrections.

Evaluation of the latter corrections is the main purpose of
the present paper. Renormalization scale dependence of the truncated
perturbation series is going to be significantly reduced. In our case, it
refers to the branching ratio dependence on the scale
$\mu_0$ at which the top-quark mass and $\alpha_s$ are renormalized.  At the
two-loop order, the corresponding uncertainty amounts to around $1.8\%$,
which is a non-negligible component of the overall theoretical uncertainty.

We introduce the effective Lagrangian as
\begin{eqnarray}
  {\mathcal L}_{\rm eff} &=& {\mathcal L}_{\rm QCD \times QED}\mbox{(leptons and five light
    quarks)} + N \sum_n C_n Q_n ~+~ {\rm h.c.}\,,
  \label{eq::leff}
\end{eqnarray}
with
\begin{eqnarray}
  N &=& \frac{V_{tb}^* V_{ts} G_F^2 M_W^2}{\pi^2}\,,
\end{eqnarray}
and the operators
\begin{eqnarray}
  Q_A &=& (\bar{b} \gamma_{\alpha} \gamma_5 s)(\bar{\mu} \gamma^{\alpha}
  \gamma_5 \mu)\,, \nonumber\\
  Q_S &=& (\bar{b} \gamma_5 s)(\bar{\mu} \mu)\,,\nonumber\\
  Q_P &=& (\bar{b} \gamma_5 s)(\bar{\mu} \gamma_5 \mu)
  \,.
\end{eqnarray}
In the SM, the operator $Q_A$ alone is sufficient because
contributions from $Q_S$ and $Q_P$ to the branching ratio are suppressed by
$M^2_{B_s}/M_W^2$ with respect to that from $Q_A$. In beyond-SM
theories, the Wilson coefficients $C_S$ and $C_P$ can get
enhanced, especially for an extended Higgs sector (see, e.g.,
Refs.~\cite{Logan:2000iv,Bobeth:2001jm}).  Note that $Q_V =
(\bar{b}\gamma_{\alpha}\gamma_5 s)(\bar{\mu} \gamma^{\alpha} \mu)$ does
not contribute at the leading order in $\alpha_{em}$ due to the 
electromagnetic current conservation.

Using Eq.~(\ref{eq::leff}), the following result for the
average time integrated branching ratio can be derived
\begin{equation}
  \overline{\mathcal B}(B_s\to \mu^+ \mu^-) = 
  \frac{|N|^2 M_{B_s}^3 f_{B_s}^2}{8 \pi\,
  \gammas_H} \beta 
   \Big[ |r C_A - u C_P|^2 F_P + |u \beta C_S|^2 F_S\Big] 
   ~+~ {\mathcal O}(\alpha_{em})\,,
 \label{eq::BR}
\end{equation}
where 
% $M_{B_s}$ is the $B_s$ meson mass, $f_{B_s}$ is the leptonic decay constant, and 
$\gammas_H$ stands
for the total width of the heavier mass eigenstate in the $B_s\bar B_s$ system.
The quantities $r$, $\beta$ and $u$ are given by
\begin{eqnarray}
  r &=& \frac{2 m_{\mu}}{M_{B_s}},\qquad 
  \beta = \sqrt{1-r^2},\qquad 
  u = \frac{M_{B_s}}{m_b + m_s} \,.
\end{eqnarray}
In the absence of beyond-SM sources of CP-violation, we have $F_P=1$ and
$F_S = 1 - \Delta \gammas/\gammas_L$, where $\gammas_L$ is the lighter
eigenstate width, and $\Delta \gammas = \gammas_L - \gammas_H$. In a generic 
case, from the results in Refs.~\cite{DeBruyn:2012wj,DeBruyn:2012wk} one derives
\begin{eqnarray}
F_P &=& 1 -\frac{\Delta\gammas}{\gammas_L} \sin^2 \left[ \frac12 \phi_s^{\rm NP} + \arg(r C_A-u C_P)\right]\,,
\nonumber\\
F_S &=& 1 -\frac{\Delta\gammas}{\gammas_L} \cos^2 \left[ \frac12 \phi_s^{\rm NP} + \arg C_S\right]\,,
\end{eqnarray}
where $\phi_s^{\rm NP}$ describes the CP-violating ``new physics''
contribution to $B_s\bar B_s$ mixing, i.e. 
$\phi_s^{c\bar c s} \simeq \arg[(V_{ts}^* V_{tb})^2] + \phi_s^{\rm NP}$
(see Sec. 2.2 of Ref.~\cite{Buras:2013uqa}).

In the SM, the branching ratio of $B_s \to \mu^+ \mu^-$ is proportional
to the square of the Wilson coefficient $C_A$ which can be computed within
perturbation theory. The calculation amounts to matching the amplitude\footnote{
More precisely, we shall match the $\bar bs\bar\mu\mu$
one-light-particle-irreducible (1LPI) Green's functions at vanishing external
momenta.}
for $s \to b \, \mu^+ \mu^-$ in the full SM to the one of the
effective theory defined in Eq.~(\ref{eq::leff}).  At the matching scale
$\mu_0$, the $W$ and $Z$ bosons together with the top quark are
integrated out simultaneously. 

Barring higher-order electroweak (EW) corrections,
the perturbative expansion of $C_A$ reads
\begin{eqnarray} \label{eq::cAexp}
  C_A &=& C_A^{(0)} + \frac{\alpha_s}{4\pi} C_A^{(1)} +
  \left(\frac{\alpha_s}{4\pi}\right)^2 C_A^{(2)} + ... \,,
\end{eqnarray}
where $\alpha_s\equiv\alpha_s(\mu_0)$ in the $\overline{\rm MS}$ scheme
with five active quark flavours. No other definition of $\alpha_s$ is going to be
used throughout the paper. The one-loop term $C_A^{(0)}$ has been
calculated for the first time in Ref.~\cite{Inami:1980fz}, and the
two-loop correction $C_A^{(1)}$ has been found in
Refs.~\cite{Buchalla:1992zm,Buchalla:1993bv,Misiak:1999yg,Buchalla:1998ba}. 
In this work, we compute the three-loop QCD correction $C_A^{(2)}$. 

Let us note that $C_A^{(n)}$ are $\mu_0$-dependent, but $C_A$ itself is not,
up to higher-order QED effects. It follows from the fact that the quark
current in $Q_A$ is classically conserved in the limit of vanishing quark
masses, while the chiral anomaly plays no role here, as we work at the leading
order in flavour-changing interactions. Once the perturbation series on the
r.h.s. of Eq.~(\ref{eq::cAexp}) is truncated, a residual $\mu_0$-dependence
arises. Our present calculation aims at making this dependence practically
negligible.

At each loop order, we shall split the 
coefficients $C_A^{(n)}$ into contributions originating from the
$W$-boson box and the $Z$-boson penguin diagrams (see
Figs.~\ref{fig::w_boxes} and~\ref{fig::z_penguin})
\begin{eqnarray}
  C_A^{(n)} &=& C_A^{W,(n)} + C_A^{Z,(n)}\,,
\end{eqnarray}
which are separately finite but gauge-dependent with respect to the
EW gauge fixing.  Here, we use the background field version of the
't~Hooft-Feynman gauge for the electroweak bosons, and the usual
't~Hooft-Feynman gauge for the gluons. Most of the results have also
been cross-checked using the general $R_\xi$ gauge for the gluons.

For the top quark mass renormalization, we shall always use the
    $\overline{\rm MS}$ scheme in the full SM, i.e. $m_t \equiv
    m_t(\mu_0)$. The ratio $m_t/M_W$ will enter our results via the
    following three variables:
\begin{align}
  &x = \frac{m_t^2}{M_W^2}\,,&&
  w = 1-\frac{1}{x}\,, && y = \frac{1}{\sqrt{x}} \,.
\end{align}
The ratio $x$ is the only parameter on which the coefficients $C_A^{(n)}$
depend, apart from the logarithms $\ln(\mu_0/M_W)$ or $\ln(\mu_0/m_t)$. 
For the $Z$-penguins, this is true after taking the leading-order EW relations
between $M_Z$, $M_W$ and $\sin^2\theta_W$ into account.

Our paper is organized as follows: in the next two sections, we
evaluate the matching coefficient $C_A$ up to three loops. 
Calculations of the $W$-boxes and the $Z$-penguins are discussed in
Sections~\ref{sec::WBox} and~\ref{sec::ZPenguin}, respectively.
Section~\ref{sec::num} is devoted to a numerical analysis and examining
the size of the evaluated three loop QCD corrections. We conclude in
Section~\ref{sec::conclusions}. Logarithmically enhanced QED corrections
to $C_A$ are summarized in the Appendix.

%- }}}

%- {{{ W boxes:

\section{\label{sec::WBox}$W$-boson boxes}

\subsection{General remarks}

\begin{figure}[t]
  \begin{center}
    \includegraphics[width=\textwidth]{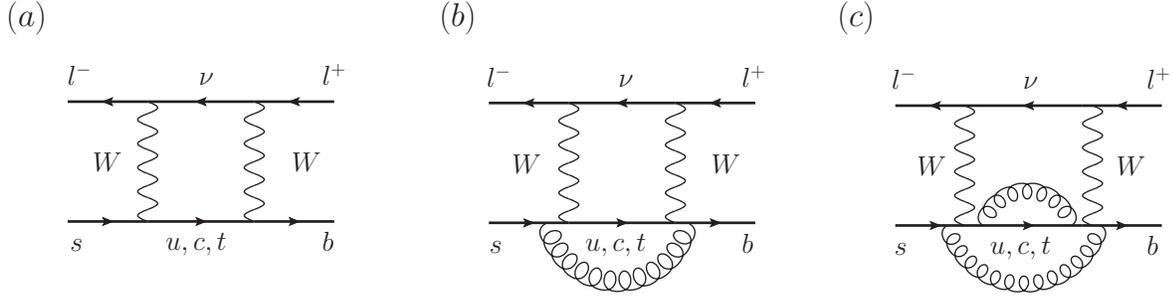}
    \caption{\label{fig::w_boxes} Sample $W$-boson box diagrams contributing to $C_A$.}
  \end{center}
\end{figure}

Sample Feynman diagrams contributing to $C_A^{W,(n)}$ at one-, two- and
three-loop order are shown in Fig.~\ref{fig::w_boxes}. The up- and
charm-quark contributions differ only by the corresponding
Cabibbo-Kobayashi-Maskawa (CKM) factors because we neglect masses of these
quarks. Consequently, it is possible to write $C_A^{W,(n)}$ in terms of the
top- and charm-quark contributions
\begin{eqnarray}
  C_A^{W,(n)} &=& C_A^{W,t,(n)} - C_A^{W,c,(n)}\,,
  \label{eq::CAW_split}
\end{eqnarray}
where unitarity of the CKM matrix has been applied.

To obtain $C_A^{W,t,(n)}$ and $C_A^{W,c,(n)}$, we compute off-shell 
1LPI amplitudes both in the full theory and in the effective
theory, and require that they agree at the scale $\mu_0$ up to terms
suppressed by heavy masses. In fact, on the full-theory side, all 
the external momenta can be set to zero, which leads to vacuum
integrals up to three loops.  On the other hand, in the effective 
theory, all loop corrections vanish in dimensional
regularization after setting the external momenta and light quark masses to
zero because the loop integrals are scaleless in this
limit. Thus, we are only left with tree contributions.

There are basically two approaches to perform the matching. In the first
one, the matching is performed in $d=4-2\epsilon$ dimensions setting
all the light masses strictly to zero. As a consequence, one 
generates spurious infrared divergences both in the full and effective
theories. Such divergences cancel while extracting $C_A$.  However, due
to the presence of additional poles in $\epsilon$ at intermediate
steps, one has to introduce the so-called evanescent operators
in the effective Lagrangian, which complicates the calculations.  In an
alternative matching procedure, finite light quark masses are
introduced to obtain infrared and ultraviolet finite results,
which allows for a matching in four dimensions. In the latter
case, no evanescent operators matter.  In the following, we
describe both matching procedures in more detail.

\subsection{Matching in $d$ dimensions}

The evanescent operator which enters the effective Lagrangian
when the matching is performed in $d$ dimensions reads~\cite{Misiak:1999yg}
\begin{eqnarray}
  Q^E_A &=& (\bar{b} \gamma_{\alpha_1} \gamma_{\alpha_2} \gamma_{\alpha_3}
  \gamma_5 s)(\bar{\mu} \gamma^{\alpha_3} \gamma^{\alpha_2} \gamma^{\alpha_1}
  \gamma_5 \mu) - 4 \, Q_A\,.
\end{eqnarray}
Note that this operator vanishes in $d=4$ dimensions, and thus the 
limit $d \to 4$ can only be taken after matching in $d$
dimensions.

Before performing the matching, we have to replace the combination $C_A
Q_A + C^E_A Q_A^E$ by the corresponding renormalized expression 
that can be written as~\cite{Misiak:1999yg}
\begin{eqnarray}
  C_A Q_A + C^E_A Q_A^E
  \to Z_{\psi} 
  \left( C_A Z_{NN} Q_A + C_A Z_{NE} Q_A^E + C^E_A Z_{EN} Q_A 
    + C^E_A Z_{EE} Q^E_A \right)
  \,,
\end{eqnarray}
where $Z_{\psi}$ is the $\overline{\rm MS}$ quark wave function renormalization
constant. Loop corrections to $Z_{\psi}$, $Z_{NN}$, $Z_{EE}$ and
$Z_{NE}$ contain no finite parts\footnote{
In our conventions, $n$-loop integrals are normalized with
$\widetilde{\mu}^{n\epsilon}$, where $\widetilde{\mu}^2 \equiv
\mu^2e^\gamma/(4\pi)$ and $\gamma$ denotes the Euler-Mascheroni
constant. Thus, no $\ln 4\pi$ or $\gamma$ appear in the $\overline{\rm MS}$
renormalization constants.}
but at most poles in $\epsilon$.  As far as $Z_{EN}$ is concerned, we require
that amplitudes proportional to $C_A^E$ vanish for $d\to4$. In consequence,
$Z_{EN}$ may contain both pole parts and (uniquely defined) finite terms. For
our purpose, the renormalization constants are needed up to two loops.
\begin{figure}[t]
  \begin{center}
      \includegraphics[width=0.6\textwidth]{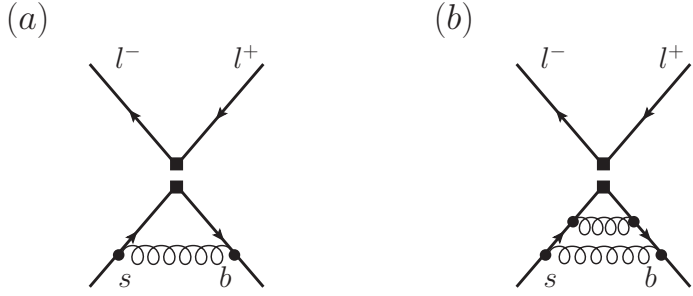}
    \caption{
      \label{fig::eft}
      Sample one- and two-loop Feynman diagrams needed for determination of
      the renormalization constants $Z_{NN}$, $Z_{NE}$, $Z_{EN}$
      and $Z_{EE}$. Squares represent the operators
      $Q_A$ and $Q_A^E$.}
  \end{center}
\end{figure}

The renormalization constants $Z_{NN}$, $Z_{NE}$, $Z_{EN}$ and $Z_{EE}$ are
computed from the diagrams like those in Fig.~\ref{fig::eft},
with insertions of $Q_A$ and $Q_A^E$. Since we are only interested 
in ultraviolet poles of momentum integrals, all the masses
can be set to zero, and an external momentum $q$ flowing through the
quark lines is introduced.  Our results read
\begin{eqnarray}
  Z_{NN} &=& 1\,, \nonumber\\
  Z_{NE} &=& 0\,, \nonumber\\
  Z_{EN} &=& \frac{\alpha_s}{4\pi}\, 32 + \left(\frac{\alpha_s}{4\pi}\right)^2
  \left[ \frac{1}{\epsilon} \left(  
      -176 + \frac{32}{3} n_f \right) + \frac{1192}{3} - \frac{112}{9} n_f
  \right] +\; {\mathcal O}(\alpha_s^3)\,, \nonumber\\
  Z_{EE} &=& 1 + {\mathcal O}(\alpha_s^2) \,,
\end{eqnarray}
where $n_f=5$ denotes the number of active quark 
flavours. The results for $Z_{NN}$ and $Z_{NE}$ are true to all
orders in QCD due to the (already mentioned) quark current
conservation for massless quarks. Concerning $Z_{EN}$, we 
confirm the one-loop result from Ref.~\cite{Misiak:1999yg},
whereas the two-loop expression is new. Note that $Z_{EE}$ does not matter
for our calculation, and thus we have left its two-loop part
unevaluated.

In the first step of our matching calculation, we
determine the $s\to b\mu^+\mu^-$ transition amplitude
in the full theory, where the Dirac structure of each Feynman
diagram is projected onto $Q_A$ and $Q_A^E$ (see, e.g.,
Ref.~\cite{Bobeth:1999mk}). This gives us the 
unrenormalized amplitudes, which we denote by
$C_{A,\text{bare}}^{W}$ and $C_{A,\text{bare}}^{E}$,
respectively.  In this step,
vacuum diagrams up to three loops with two different mass scales have to be
computed. Although some classes of Feynman diagrams of this type have been
studied in the literature (see, e.g., Ref.~\cite{Grigo:2012ji}), we
have decided to perform expansions in various limits, which leads to
handy results for the matching coefficients. Actually, we follow the same
strategy as in Refs.~\cite{Misiak:2004ew,Hermann:2012fc}, namely, we
expand in the limits $M_W\ll m_t$ and $M_W\approx m_t$, i.e. $y\ll 1$ and
$w\ll 1$, where terms up to order $y^{12}$ and $w^{16}$ are evaluated,
respectively.  A simple combination of the two expansions provides an
approximation to the three-loop contribution, which for all
practical purposes is equivalent to an exact result.

The actual calculation has been performed with the help of {\tt
  QGRAF}~\cite{Nogueira:1993ex} to generate the Feynman diagrams, {\tt q2e}
  and {\tt exp}~\cite{q2eexp} for the asymptotic
  expansions~\cite{Smirnov:2012gma} and {\tt MATAD}~\cite{Steinhauser:2000ry},
  written in {\tt Form}~\cite{Form}, for evaluation of the three-loop
  diagrams. We have performed our calculation for an 
  arbitrary gauge parameter in QCD, and have checked that it drops out in
  our final result for the matching coefficient.

For renormalization of the full-theory contributions, we need
the one-loop renormalization constant for the QCD gauge coupling
\begin{equation}
  Z_g^{\rm SM} = 1+ \frac{\alpha_s}{4\pi} 
  \left( -\frac{23}{6\epsilon} + \frac{1}{3\epsilon} N_{\epsilon} \right) + {\mathcal O}(\alpha_s^2)
  \,.
\end{equation}
Here,~ $N_{\epsilon} = ( \mu_0^2/m_t^2 )^{\epsilon}\, e^{\gamma \epsilon}\, \Gamma(1 + \epsilon)$
makes the renormalized $\alpha_s$ in the full SM equal to the $\overline{\rm
MS}$-renormalized $\alpha_s$ in the five-flavour effective theory, to all
orders in $\epsilon$. As far as the top quark mass is concerned, its two-loop
$\overline{\rm MS}$ renormalization constant $Z_{m_t}$ in the full SM is expressed in 
terms of the above-defined $\alpha_s$, which gives
\begin{equation}
Z_{m_t} = 1 -\frac{4}{\epsilon} \frac{\alpha_s}{4\pi} + \left(\frac{\alpha_s}{4\pi}\right)^2
\left( \frac{74}{3\epsilon^2} - \frac{27}{\epsilon} - \frac{8}{3\epsilon^2} N_{\epsilon} \right) 
+ {\cal O}(\alpha_s^3).
\end{equation}
Furthermore, for the wave-function renormalization, only the difference
between the renormalization constants in the full and effective
theories has to be taken into account (see Section~4 of
Ref.~\cite{Misiak:2004ew}):
\begin{equation}
  \Delta Z_{\psi} =
  \left(\frac{\alpha_s}{4\pi}\right)^2 N_{\epsilon}^2 
  \left( \frac{2}{3\epsilon} - \frac{5}{9} \right) + {\mathcal O}(\alpha_s^3,\epsilon)
  \,.
\end{equation}

At this point, all the ingredients are available to perform the matching
according to the following equations ($Q=c,t$):
\begin{eqnarray}
  C_{A}^{E,Q} &=& C_{A,\text{bare}}^{E,Q,(0)}
  + \frac{\alpha_s}{4 \pi} \left( C_{A,\text{bare}}^{E,Q,(1)}
    + \delta^{t Q} \Delta T^{E,t,(1)} \right) + {\mathcal O}\left(\alpha_s^2\right)
  \,,
  \nonumber\\
  C_A^{W, Q} &=& \left(1 + \Delta Z_{\psi} \right) \sum\limits_{n=0}^2 \left(
    \frac{\alpha_s}{4 \pi} \right)^n \left[ \left( Z_g^{\text{SM}} \right)^{2
      n} C_{A,\text{bare}}^{W,Q,(n)} 
    + \delta^{t Q} \Delta T^{W,t,(n)} \right] \nonumber \\
  && - Z_{EN} \, C_{A}^{E,Q} + {\mathcal O}\left(\alpha_s^3\right)
  \,,
  \label{eq::Wmatch}
\end{eqnarray}
where $\Delta T^{E,t,(1)}$ and $\Delta T^{W,t,(n)}$ denote contributions from
the top-quark mass counter\-terms which can be written as
\begin{eqnarray}
  \Delta T^{E,t,(1)} &=& \left( C_{A,\text{bare}}^{E,t,(0)}
    \big|_{m_{t}^{\rm bare} \to Z_{m_t} m_t} \right)_{\alpha_s} 
  \,,\nonumber\\
  \Delta T^{W,t,(0)} &=& 0\,, \nonumber \\
  \Delta T^{W,t,(1)} &=& \left( C_{A,\text{bare}}^{W,t,(0)} \big|_{m_{t}^{\rm bare}
      \to Z_{m_t} m_t} \right)_{\alpha_s}\,, \nonumber \\ 
  \Delta T^{W,t,(2)} &=& \left( C_{A,\text{bare}}^{W,t,(0)} \big|_{m_{t}^{\rm bare}
      \to Z_{m_t} m_t} 
    + \frac{\alpha_s}{4\pi} C_{A,\text{bare}}^{W,t,(1)} \big|_{m_{t}^{\rm bare} \to Z_{m_t} m_t}
  \right)_{\alpha_s^2} 
  \,.
\end{eqnarray}
Here, the following notation has been used: ``$m_{t}^{\rm
bare}\to Z_{m_t} m_t$'' means that the bare top quark mass is replaced
by the renormalized one times the renormalization
constant. Afterwards, we expand in $\alpha_s$ and take the coefficient
at $[\alpha_s/(4\pi)]^n$ ($n=1,2$), which is indicated by the
subscript at the round bracket.

Our final results for the evanescent Wilson coefficients up to two
loops read
\begin{eqnarray}
  C_{A}^{E,t,(0)} &=& \frac{1}{64} \left(\frac{\mu_0^2}{m_t^2}\right)^\epsilon 
   \left[ \frac{2}{x-1} - \frac{2x\ln x}{(x-1)^2}  + \epsilon 
    \left( \frac{3}{x-1} - \frac{(x+2) \ln x + x \ln^2 x}{(x-1)^2} \right)\right] 
    + {\mathcal O}(\epsilon^2)
  \,,\nonumber\\[2mm]
  C_{A}^{E,t,(1)} &=& \frac{7-23x}{24(x-1)^2} + \frac{7x+9 x^2}{24(x-1)^3} \ln x 
  +\frac{x}{4 (x-1)^2} \,\mbox{Li}_2\left(1-\frac{1}{x}\right) \nonumber \\[1mm]
  && + \ln\left(\frac{\mu_0^2}{m_t^2}\right) \left[ \frac{-x}{2(x-1)^2} + 
   \frac{x + x^2 }{4(x-1)^3} \ln x \right] ~+~ {\mathcal O}(\epsilon)
  \,,\nonumber\\[2mm]
  C_{A}^{E,c,(0)} &=& -\frac{1}{64} \left(\frac{\mu_0^2}{M_W^2}\right)^\epsilon 
                      (  2  + 3\epsilon ) ~+~ {\mathcal O}(\epsilon^2)
  \,,\nonumber\\[2mm]
  C_{A}^{E,c,(1)} &=& \frac{7}{24} ~+~ {\mathcal O}(\epsilon)
  \,.
\end{eqnarray}
The results for $C_A^{W,t}$ and $C_A^{W,c}$ will be given in
Subsection~\ref{subsec::results}. 

\subsection{Matching in four dimensions}

In order to have a cross check of the results for $C_A^W$ from the previous 
subsection, we have performed the matching also for infrared finite 
quantities, which can be done in four dimensions avoiding evanescent
operators~\cite{Misiak:1999yg}. No spurious infrared divergences 
arise when small but non-vanishing masses are introduced for the strange
and bottom quarks. In the full theory, this leads to Feynman diagrams with
up to four different mass scales. We evaluate them using asymptotic 
expansions in the limit
\begin{equation}
  m_t, M_W \gg m_b \gg m_s\,.
\end{equation}
In addition, we use either $m_t \gg M_W$ or $m_t \approx M_W$, as in the
previous subsection. All the external momenta are still set to zero. Asymptotic
expansions are conveniently performed with the help of {\tt exp}~\cite{q2eexp}.

On the effective-theory side, the loop corrections do not vanish any more due
to the finite quark masses. We compute the necessary one- and two-loop 
Feynman integrals in the limit
\begin{equation}
 m_b \gg m_s\,.
\end{equation}

After renormalization of the two-loop expression on the effective-theory side
and the three-loop result on the full-theory side, the finite parts are
matched for $\epsilon\to 0$. After the matching, it is possible to take
the limit $m_s \to 0$ and $m_b \to 0$. This way, we
obtain the same results as in the previous calculation where the
infrared divergences have been regulated using dimensional regularization.

Although we only had to compute the leading non-vanishing contributions
in the light quark masses, the calculational effort has been
significantly higher than for the matching in $d$ dimensions described in the
previous subsection.  Thus, we have applied the method with light
masses only to cross check the first two (three) terms in the expansion in
$y$ ($w$), using a general $R_\xi$ gauge though.

\subsection{\label{subsec::results}Results}

At the one- and two-loop orders, we have confirmed the results with
full dependence on $x$ from Ref.~\cite{Misiak:1999yg}, and evaluated in
addition terms up to ${\mathcal O}(\epsilon^2)$ and ${\mathcal O}(\epsilon)$,
respectively. For completeness, we present the results for $\epsilon\to 0$
which are given by
\begin{eqnarray}
  C_A^{W,t,(0)}(\mu_0) &=& \frac{1}{8(x-1)} - \frac{x}{8 (x-1)^2} \ln x
  \,,
  \nonumber\\[2mm]
  C_A^{W,t,(1)}(\mu_0) &=&
  -\frac{3 + 13 x}{6 (x-1)^2} +\frac{17 x-x^2}{6 (x-1)^3} \ln x
 +\frac{x}{(x-1)^2} \,\mbox{Li}_2\left(1-\frac{1}{x}\right) \nonumber\\[1mm]
  && + \ln\left(\frac{\mu_0^2}{m_t^2}\right) \left[ \frac{-2x}{(x-1)^2}
      + \frac{x+x^2}{(x-1)^3} \ln x \right]
  \,,
  \nonumber\\[2mm]
  C_A^{W,c,(0)}(\mu_0) &=& - \frac{1}{8}
  \,,
  \nonumber\\[2mm]
  C_A^{W,c,(1)}(\mu_0) &=& - \frac{1}{2}
  \,.
\end{eqnarray}
Analytic expressions including ${\mathcal O}(\epsilon)$ terms can be
downloaded from~\cite{progdata}.

With the help of the exact two-loop result, we can extract the full 
$x$-dependence in front of the $\ln \mu_0$ terms at the three-loop level. We
find
\begin{eqnarray}
  C_A^{W,t,(2)}(\mu_0) &=& C_A^{W,t,(2)}(\mu_0 = m_t)
     + \ln\left(\frac{\mu_0^2}{m_t^2}\right) \left[ 
     \frac{69 +1292 x-209 x^2}{18 (x-1)^3} \right. \nonumber \\[2mm]
  && \left. - \frac{ 521 x + 105 x^2 - 50 x^3 }{9(x-1)^4} \ln x
    -\frac{47 x+x^2}{3(x-1)^3} \,\mbox{Li}_2\left(1-\frac{1}{x}\right)\right] \nonumber \\[2mm]
  && + \ln^2 \left(\frac{\mu_0^2}{m_t^2}\right) \left[ \frac{61x+11x^2}{3(x-1)^3}  
     - \frac{49x+96x^2-x^3}{6 (x-1)^4} \ln x \right]
  \,,  \nonumber\\[2mm]
  C_A^{W,c,(2)}(\mu_0) &=& C_A^{W,c,(2)}(\mu_0 = M_W) -\frac{23}{6} \ln \left(\frac{\mu_0 ^2}{M_W^2}\right)
  \,.
\end{eqnarray}
We have chosen $\mu_0 = m_t$ and $\mu_0 = M_W$ as default scales for the top
and charm sectors, respectively.  Analytical results for all the 
coefficients can be downloaded from~\cite{progdata}. In the following, we
present the results in a compact numerical form.  For our two expansions, the
coefficient in the charm sector reads
\begin{eqnarray}
  C_A^{W,c,(2)}(\mu_0 = M_W) &=& 
- 5.222
- 0.2215 \, y^2
+ 0.1244 \, y^2 \ln y
- 0.08889 \, y^2 \ln^2 y
+ 0.04146 \, y^4 \nonumber \\
&& - 0.02955 \, y^4 \ln y
+ 0.009524 \, y^4 \ln^2 y
- 0.001092 \, y^6
+ 0.0006349 \, y^6 \ln y \nonumber \\
&& - 0.00004286 \, y^8
+ 0.00003207 \, y^8 \ln y
- 3.109 \cdot 10^{-6} \, y^{10} \nonumber \\
&& + 2.643 \cdot 10^{-6} \, y^{10} \ln y 
- 3.009 \cdot 10^{-7} \, y^{12}
+ 2.775 \cdot 10^{-7} \, y^{12} \ln y \nonumber \\
&& + {\mathcal O}\left(y^{14}\right)
  \,,   \label{eq::res_W_c_a} \\
  C_A^{W,c,(2)}(\mu_0 = M_W) &=&
- 5.403
+ 0.09422 \, w
+ 0.02786 \, w^2
+ 0.01355 \, w^3
+ 0.008129 \, w^4 \nonumber \\
&& + 0.005469 \, w^5
+ 0.003957 \, w^6
+ 0.003009 \, w^7
+ 0.002373 \, w^8 \nonumber \\
&& + 0.001925 \, w^9 
+ 0.001596 \, w^{10}
+ 0.001346 \, w^{11}
+ 0.001153 \, w^{12} \nonumber \\
&& + 0.0009996 \, w^{13} 
+ 0.0008757 \, w^{14}
+ 0.0007742 \, w^{15}
+ 0.0006898 \, w^{16} \nonumber \\
&& + {\mathcal O}\left(w^{17}\right)
  \,.
  \label{eq::res_W_c}
\end{eqnarray}
The corresponding coefficient in the top sector is given by
\begin{eqnarray}
  C_A^{W,t,(2)}(\mu_0 = m_t) &=&
2.710 \, y^2
+ 6.010 \, y^2 \ln y
- 8.156 \, y^4
- 1.131 \, y^4 \ln y
- 0.5394 \, y^6 \nonumber \\
&& - 13.97 \, y^6 \ln y
+ 35.32 \, y^8
+ 15.64 \, y^8 \ln y
+ 103.9 \, y^{10} \nonumber \\
&& + 149.2 \, y^{10} \ln y
+ 207.7 \, y^{12}
+ 454.8 \, y^{12} \ln y
+ {\mathcal O}\left(y^{14}\right)
  \,, \label{eq::res_W_t_a} \\
  C_A^{W,t,(2)}(\mu_0 = m_t) &=&
- 0.4495
- 0.5845 \, w
+ 0.1330 \, w^2
+ 0.1563 \, w^3
+ 0.1233 \, w^4 \nonumber \\
&& + 0.09333 \, w^5
+ 0.07134 \, w^6
+ 0.05561 \, w^7
+ 0.04425 \, w^8 \nonumber \\
&& + 0.03589 \, w^9
+ 0.02960 \, w^{10}
+ 0.02478 \, w^{11}
+ 0.02102 \, w^{12} \nonumber \\
&& + 0.01803 \, w^{13}
+ 0.01562 \, w^{14}
+ 0.01366 \, w^{15}
+ 0.01204 \, w^{16} \nonumber \\
&& + {\mathcal O}\left(w^{17}\right)
  \,.
  \label{eq::res_W_t}
\end{eqnarray}
\begin{figure}[t]
  \begin{center}
    \begin{tabular}{cc}
      \includegraphics[width=0.48\textwidth]{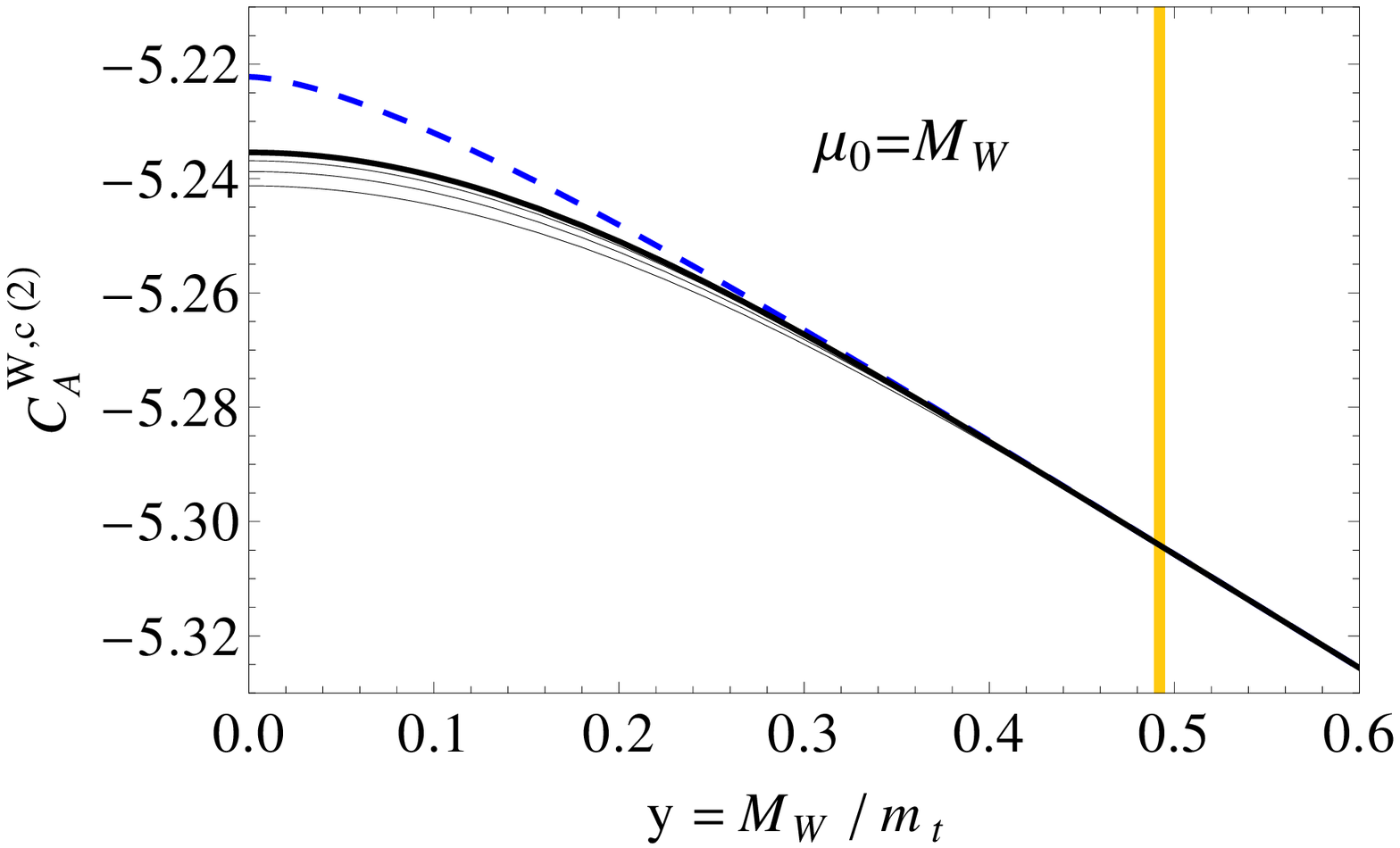} &
      \includegraphics[width=0.48\textwidth]{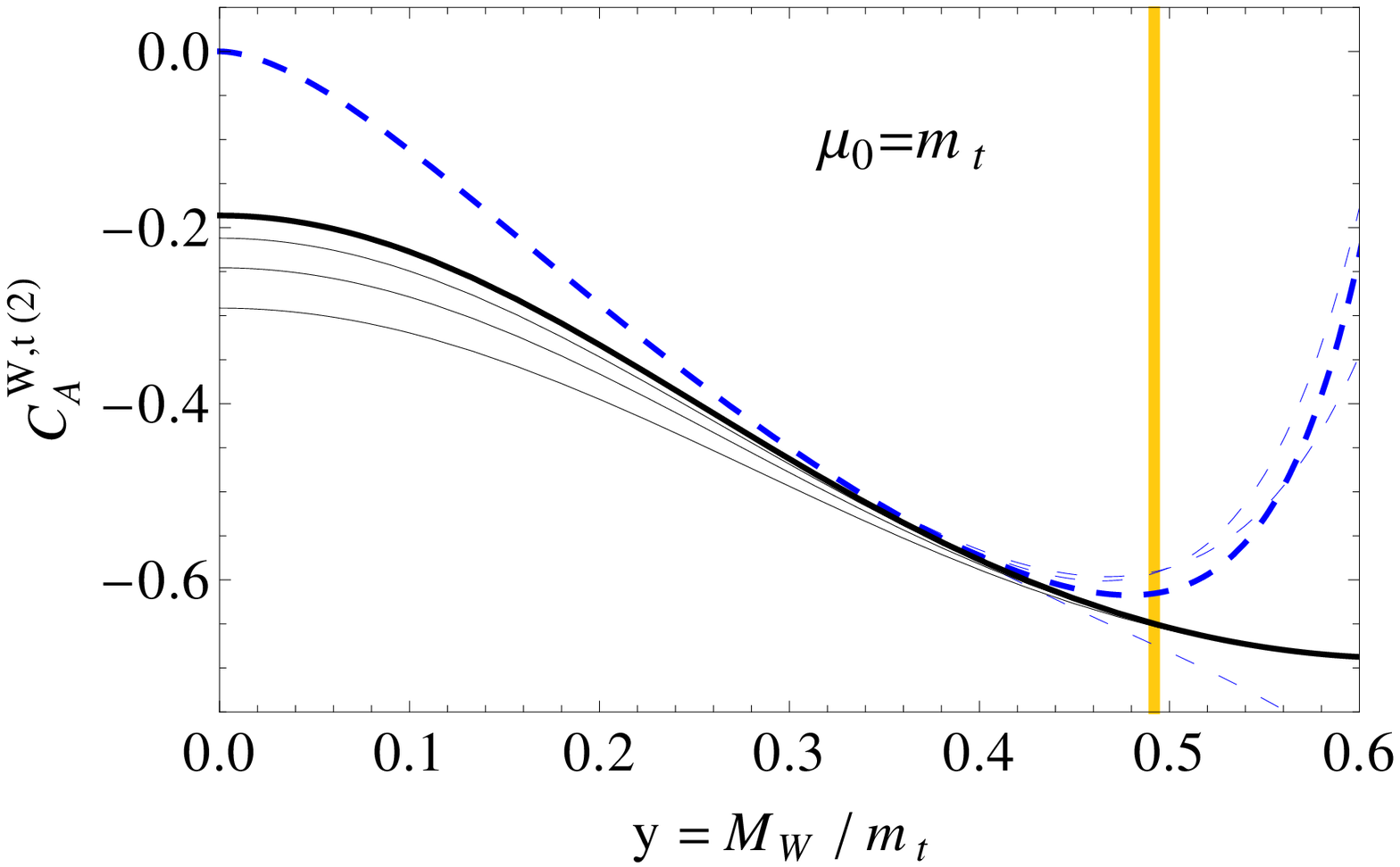}
    \end{tabular}
     \caption{
      \label{fig::result_W_box}
      $C_A^{W,(2)}$ as a function of $y=M_W/m_t$ for the charm (left) and
      top quark sector (right). The (blue) dashed lines are obtained in the 
      limit $y\ll 1$, and the (grey) solid line for $w=1-y^2\ll 1$.
      Thinner lines contain less terms in the expansions. The physical region
      for $y$ is indicated by the (yellow) vertical band.}
  \end{center}
\end{figure}

In Fig.~\ref{fig::result_W_box}, the results from
Eqs.~(\ref{eq::res_W_c_a})--(\ref{eq::res_W_t}) are shown as
functions of $y=M_W/m_t$.  The dashed and solid lines
correspond to the $y\to 0$ and $y=\sqrt{1-w}\to 1$ expansions,
respectively. Thin lines are obtained by using less expansion terms in $y$ and
$w$. They can be used to test convergence of the expansions, as it is
expected that good agreement with the unknown exact result is achieved 
up to the point where two successive orders almost coincide.

In the case of $C_A^{W,c,(2)}$ (left panel of Fig.~\ref{fig::result_W_box}), 
there is a significant overlap of the expansions around the two limits
in the region from $y\approx 0.3$ to $y\approx 1.4$. The agreement over
such a large range arises probably due to the relatively simple
dependence of $C_A^{W,c,(2)}$ on the top quark mass: $m_t$ only occurs through
one-loop corrections to the gluon propagator.  Note also that the numerical
effect of the top quark mass is moderate: $C_A^{W,c,(2)}$ changes
only by around 1.5\% between the $m_t\to\infty$ limit and the
physical value of $m_t$.

Also in the case of $C_A^{W,t,(2)}$ (right panel of
Fig.~\ref{fig::result_W_box}) one observes an overlap of the expansions for
$y\to 0$ and $y\to 1$ around $y\approx 0.35$. This feature allows us to use
the expression in Eq.~(\ref{eq::res_W_t_a}) for $y\le 0.35$, and the one in
Eq.~(\ref{eq::res_W_t}) for $y>0.35$.  Due to the convergence properties
(cf. thin lines) these expressions are excellent approximations
to the exact result in the respective regions.  In particular,
for the physical region of $y$, it is sufficient to use the expansion around
$m_t=M_W$.

For practical applications, it is useful to have short formulae which
approximate $C_A$ near the physical value of $y$. In the
range $0.3 < y < 0.7$, the fits
\begin{eqnarray}
 C_A^{W,c, (2)}(\mu_0 = M_W) &\simeq& -0.015 y^2 - 0.182 y - 5.211  \,,\nonumber\\
 C_A^{W,t, (2)}(\mu_0 = m_t) &\simeq& 2.255 \, y^2 - 2.816 \, y + 0.189\,
\end{eqnarray}
are accurate to better than 1\% in the corresponding quantities.

%- }}}

%- {{{ Z penguins:

\section{\label{sec::ZPenguin}$Z$-boson penguins}

\begin{figure}[t]
  \begin{center}
    \includegraphics[width=\textwidth]{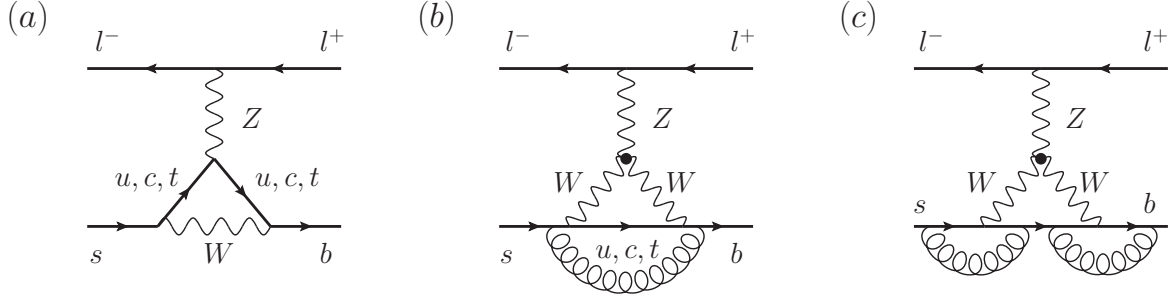}
    \caption{\label{fig::z_penguin} Sample $Z$-boson penguin diagrams contributing to $C_A$.}
  \end{center}
\end{figure}

\subsection{General remarks}

The second type of contribution to $C_A$ arises from the so-called
$Z$-boson penguins.  Sample diagrams at the one-, two- and three-loop
orders are shown in Fig.~\ref{fig::z_penguin}.  In contrast to the
$W$-box diagrams, there is no contribution from evanescent operators to
$C_A^{Z,(n)}$. However, flavour non-diagonal loop contributions to the
light quark kinetic terms require introduction of an EW
counterterm which already appears at the one-loop level. The corresponding
counterterm Lagrangian (see Eq.~(13) of Ref.~\cite{Hermann:2012fc}\footnote{
Flavour non-diagonal renormalization of the mass terms does not matter 
in the present calculation because we can treat the bottom quark as massless.}) 
can be written in the following form
\begin{equation} 
  {\mathcal L}^{\rm ew}_{\rm counter} = i \frac{G_F M_W^2}{4\sqrt{2}\,\pi^2}\, 
  V^*_{tb} V_{ts} \left( Z_{2,sb}^t  - Z_{2,sb}^c  \right)\, \bar{b}_L\! \not\!\!D s_L
\,, \label{eq::zsb}
\end{equation}
where $D^\mu$ is the covariant derivative involving the neutral gauge
boson fields ($Z$, $\gamma$, $g$). While only the one-loop contributions to
$Z_{2,sb}^{Q}$ were needed in the $\bar{B}\to X_s \gamma$
case~\cite{Misiak:2004ew}, now also the two- and three-loop corrections of
order $\alpha_s$ and $\alpha_s^2$ matter. 
The two-loop ones were also necessary in Refs.~\cite{Buchalla:1992zm,Buchalla:1993bv,Misiak:1999yg,Buchalla:1998ba,Bobeth:1999mk}.
Perturbative expansions of $Z_{2,sb}^{Q}$ are conveniently written as
\begin{eqnarray}
Z_{2,sb}^{c} &=& \sum_{n=0} 
\left(\frac{\mu_0^2}{M_W^2}\right)^{(n+1)\epsilon} 
\left(\frac{\alpha_s}{4\pi}\right)^n Z_{2,sb}^{c,(n)}\,,\nonumber\\[1mm]
Z_{2,sb}^{t} &=& \sum_{n=0} 
\left(\frac{\mu_0^2}{m_t^2}\right)^{(n+1)\epsilon} 
\left(\frac{\alpha_s}{4\pi}\right)^n Z_{2,sb}^{t,(n)}\,.
\end{eqnarray}
\begin{figure}[t]
  \begin{center}
    \includegraphics[width=\textwidth]{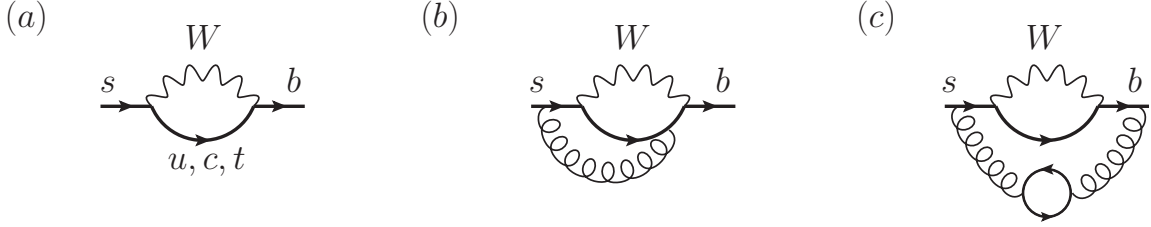}
    \caption{\label{fig::ew_ct} Sample Feynman diagrams contributing to $Z_{2,sb}^Q$.}
  \end{center}
\end{figure}
For determination of $Z_{2,sb}^{Q}$, a two-point function with incoming
strange quark and outgoing bottom quark has to be considered. Sample diagrams
at one, two and three loops are shown in Fig.~\ref{fig::ew_ct}.  We refrain
from explicitly listing the results but refer to~\cite{progdata} for
computer-readable expressions.
\begin{figure}[t]
  \begin{center}
    \includegraphics[width=\textwidth]{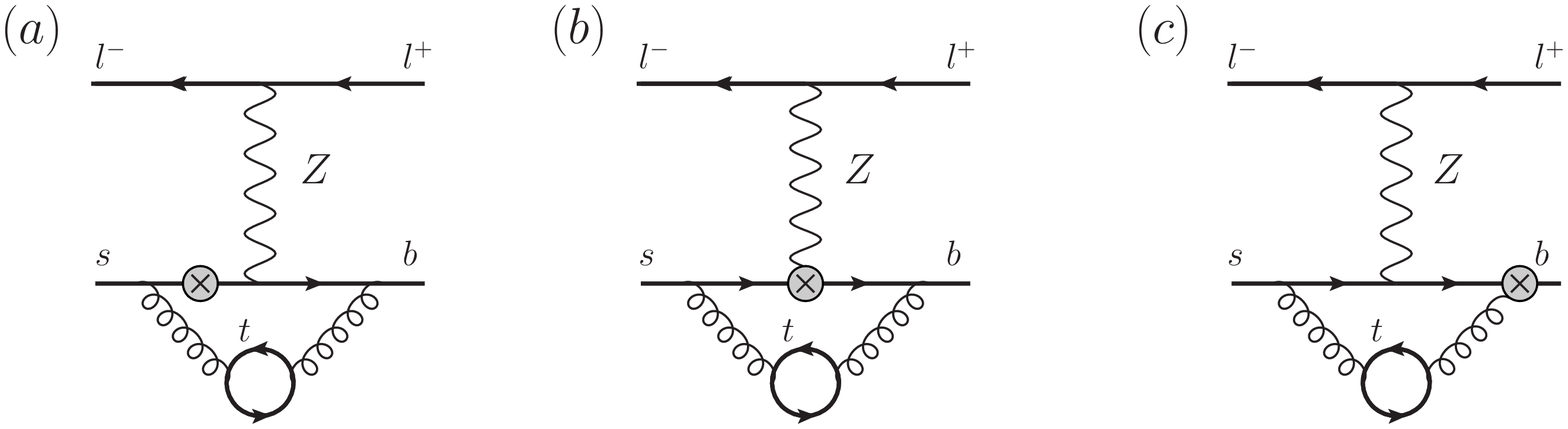}
    \caption{\label{fig::CT} Sample two-loop counterterm diagrams to the $Z$-penguin
      contribution. Altogether, there are five such diagrams.}
  \end{center}
\end{figure}

The counterterm $Z_{2,sb}^{Q}$ is either inserted in the tree-level amplitude
or in two-loop diagrams containing a closed top quark loop on the gluon
propagator, as shown in Fig.~\ref{fig::CT}. Insertions of the counterterm into other
loop contributions lead to massless tadpoles which vanish
in dimensional regularization.

\subsection{Fermion triangle contribution}

\begin{figure}[t]
  \begin{center}
    \includegraphics[width=0.7\textwidth]{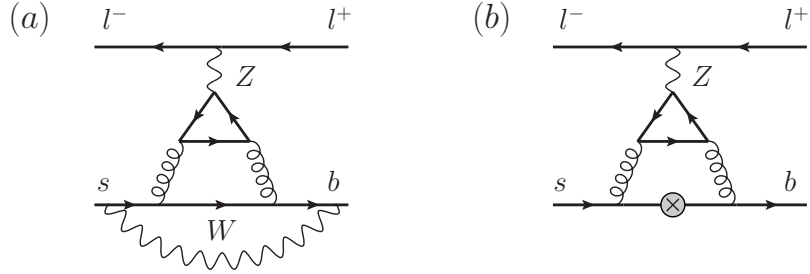}
    \caption{\label{fig::g5} Sample Feynman diagrams containing a closed triangle fermion loop
    that contribute to $C_A^{Z,(2)}$. The counterterm contribution in the right diagram 
    comes from Eq.~(\ref{eq::zsb}).}
  \end{center}
\end{figure}

There is a class of Feynman diagrams which require special attention, namely
those containing a closed triangle quark loop (see Fig.~\ref{fig::g5}).
For these contributions, a naive treatment of $\gamma_5$ as
anticommuting is not possible, and a more careful
investigation is necessary. We have followed two approaches which are
described below. Similarly to the anomaly cancellation in the
SM, contributions with the up, down, strange and charm quarks
running in the triangle loop cancel pairwise within each family. 
Thus, only the top and bottom quarks need to be considered,
as the top is the only massive quark in our calculation.

In our first approach, we adopt the prescription from Ref.~\cite{Larin:1993tq}
and replace the axial-vector coupling in the triangle loop as follows
\begin{eqnarray}
  \gamma^{\mu} \gamma_5 &\to& \frac{i}{12}\, 
  \varepsilon^{\mu\nu\rho\sigma}  
  \left( \gamma_{\nu} \gamma_{\rho} \gamma_{\sigma} - \gamma_{\sigma}
    \gamma_{\rho} \gamma_{\nu} \right) \,. 
  \label{eq::g5rep}
\end{eqnarray}
In a next step, we pull out the $\varepsilon$ tensor and 
take the trace of the loop diagram in $d$ dimensions.
In the resulting object, we perform the replacements
\begin{eqnarray}
  i\,\varepsilon^{\mu\nu\rho\sigma} \gamma_{\nu}\gamma_{\rho}\gamma_{\sigma} \gamma_5
  \otimes \gamma_\mu \gamma_5
  &\to& 6\, \gamma^\mu \otimes \gamma_\mu \gamma_5 
  \nonumber\\
  i\,\varepsilon^{\mu\nu\rho\sigma} \gamma_{\nu}\gamma_{\rho}\gamma_{\sigma}
  \otimes \gamma_\mu \gamma_5
  &\to& 6\, \gamma^\mu  \gamma_5 \otimes \gamma_\mu \gamma_5 
  \,,
\end{eqnarray}
and proceed from now on in the same way as
with the other diagrams contributing to $C_A^Z$.

In the second approach, we do not take the trace in the triangle loop
at all, but only use the cyclicity property for traces and
anticommutation relations for the $\gamma$ matrices (not for $\gamma_5$) in
order to put $\gamma_5$ to the end of each product under the trace.
Afterwards, we perform the tensor loop integration, and use
again anticommutation relations to bring the resulting expressions to	
the form
\begin{eqnarray}
  \gamma_\nu \gamma_\rho \gamma_\sigma \gamma_5 \otimes \gamma_\mu
  \gamma_{5} 
  \, \, \mbox{Tr}\left( \gamma^\nu \gamma^\rho \gamma^\sigma
    \gamma^\mu \gamma_{5} \right)
  \,,\nonumber\\
  \gamma_\nu \gamma_\rho \gamma_\sigma \otimes \gamma_\mu
  \gamma_{5} \, \, \mbox{Tr}\left( \gamma^\nu \gamma^\rho
    \gamma^\sigma \gamma^\mu \gamma_{5} \right)
  \,,
  \label{eq::stru}
\end{eqnarray}
where only the axial-vector part of the ($Z$ boson)-lepton coupling has been
taken into account.  In a next step, we add and subtract $24 \gamma^\mu \otimes
\gamma_\mu \gamma_{5}$ to the first, and $24 \gamma^\mu \gamma_5
\otimes \gamma_\mu \gamma_{5}$ to the second structure in Eq.~(\ref{eq::stru}). 
This way, we obtain the Wilson coefficients for the trace evanescent
operators~\cite{Gorbahn:2004my}
\begin{eqnarray}
  Q^E_1 &=& \gamma_\nu \gamma_\rho \gamma_\sigma \gamma_5 \otimes
  \gamma_\mu \gamma_{5} \, \, \mbox{Tr}\left( \gamma^\nu
    \gamma^\rho \gamma^\sigma \gamma^\mu \gamma_{5} \right) 
  + 24 \, \gamma^\mu \otimes \gamma_\mu \gamma_{5}
  \,,\nonumber\\
  Q^E_2 &=& \gamma_\nu \gamma_\rho \gamma_\sigma \otimes \gamma_\mu
  \gamma_{5} \, \, \mbox{Tr}\left( \gamma^\nu \gamma^\rho
    \gamma^\sigma \gamma^\mu \gamma_{5} \right) 
  + 24 \, \gamma^\mu \gamma_5 \otimes \gamma_\mu \gamma_{5}  
\end{eqnarray}
and a contribution to $C_A$. Actually, the latter is given by $(-24)$
times the prefactor of the second structure in Eq.~(\ref{eq::stru}).

The two methods, which lead to identical results for $C_A^Z$, 
have been applied both to the three-loop diagrams themselves and to
the counterterm contributions (cf. Fig.~\ref{fig::g5}).

\subsection{Matching formula}

In analogy to Eq.~(\ref{eq::CAW_split}) we can write
\begin{eqnarray}
  C_A^{Z,(n)} &=& C_A^{Z,t,(n)} - C_A^{Z,c,(n)} + \delta^{n,2} \left( C_A^{Z,t,\rm tria.} - C_A^{Z,c,\rm tria.} \right)
  \,,
\end{eqnarray}
where $C_A^{Z,Q,\rm tria.}$ are the contributions from the triangle diagrams described in the previous subsection.

The calculation of $C_A^{Z,(n)}$ proceeds along the same lines as for the
$W$-box contribution. In particular, we set all the external momenta
to zero, and expand the Feynman integrals in the full theory both for $m_t\gg
M_W$ and $m_t\approx M_W$.  Furthermore, we renormalize the top-quark mass,
$\alpha_s$ and the wave function in analogy to the $W$-box case.  As before,
all loop corrections vanish in the effective theory, which finally leads to the
following matching equation for $C_A^{Z, Q}$ ($Q=c,t$)
\begin{eqnarray}
  C_A^{Z, Q} &=& \left(1 + \Delta Z_{\psi} \right) \sum\limits_{n=0}^2 \left(
    \frac{\alpha_s}{4 \pi} \right)^n \left[ \left( Z_g^{\text{SM}} \right)^{2
      n} C_{A,\text{bare}}^{Z,Q,(n)} 
    + \delta^{t Q} \Delta T^{Z,t,(n)}  + K^{Q,(n)} \right] \nonumber\\
  &&+ \tilde{K}^{Q} + {\mathcal O}\left(\alpha_s^3\right)
  \,,
  \label{eq::CAZt_match}
\end{eqnarray}
with top-quark mass counterterms
\begin{eqnarray}
  \Delta T^{Z,t,(0)} &=& 0\,, \nonumber \\
  \Delta T^{Z,t,(1)} &=& \left( C_{A,\text{bare}}^{Z,t,(0)} \big|_{m_{t}^{\rm bare}
      \to Z_{m_t} m_t} \right)_{\alpha_s}\,, \nonumber \\ 
  \Delta T^{Z,t,(2)} &=& \left( C_{A,\text{bare}}^{Z,t,(0)} \big|_{m_{t}^{\rm bare}
      \to Z_{m_t} m_t} 
    + \frac{\alpha_s}{4\pi} C_{A,\text{bare}}^{Z,t,(1)} \big|_{m_{t}^{\rm bare}
      \to Z_{m_t} m_t} 
  \right)_{\alpha_s^2}\,.
\end{eqnarray}
$K^{Q,(n)}$ denote tree-level contributions from the
EW counterterm (\ref{eq::zsb}) which take a simple form
\begin{eqnarray}
 K^{t,(n)} &=& \left( -\frac{1}{16} + \frac{\sin^2{\theta_W}}{24}  \right)
 \left(\frac{\mu_0^2}{m_t^2}\right)^{(n+1)\epsilon} Z_{2,sb}^{t,(n)}\,,\nonumber\\[1mm]
 K^{c,(n)} &=& \left( -\frac{1}{16} + \frac{\sin^2{\theta_W}}{24}  \right)
 \left(\frac{\mu_0^2}{M_W^2}\right)^{(n+1)\epsilon} Z_{2,sb}^{c,(n)}\,.
\end{eqnarray}
Finally, $\tilde{K}^{Q}$ stands for the counterterm
contributions from two-loop diagrams like those in
Fig.~\ref{fig::CT}.  

We observe that after inserting explicit results on the right-hand side of
Eq.~(\ref{eq::CAZt_match}), all the terms proportional to
$\sin^2 \theta_W$ cancel out, and $C_A^{Z, t}$ becomes independent of
the weak mixing angle. This can be understood by recalling similarities
between the $Z$ boson and the photon couplings to other particles in the
background field gauge, as well as the structure of the counterterm in
Eq.~(\ref{eq::zsb}).  Once the quark kinetic terms in the effective theory are
imposed to be flavour diagonal, the same must be true for dimension-four
quark-photon couplings. In effect, the counterterm in Eq.~(\ref{eq::zsb})
automatically renormalizes away all the zero-momentum quark-($Z$ boson) 
interactions that come with $\sin^2 \theta_W$.

Another interesting thing to note is that $C_A^{Z,c,(n)}
= 0$ at each loop order in the background field version of the
't~Hooft-Feynman gauge, which means that only the triangle contributions
are non-vanishing in the charm sector. One of the ways to understand this
fact is again by considering diagrams where the $Z$ boson (together with
the muons) is replaced by an external photon.
%
% Let us give an argument that works to all orders in QCD.
% It is enough to consider only left-handed quarks in the flavour-changing loop, because
% only such quarks are present there in the massless (charm) case. Then the couplings of the photon
% are proportional to those of the Z boson, but with different proportionality constants.
% Let A, B, and C stand for contributions from the diagrams where the external boson (Z or gamma)
% couples to the up-quarks, down-quarks and W bosons, respectively. There are no pseudogoldstones
% in the game for massless quarks. For the photons, the sum of the diagrams gives:
%
% X_gamma = A Qu + B Qd + C = Qd (B - C) + A + C.
%
% because Qu = 1 - Qd. This means that A + C = 0, because everything must
% get renormalized away by the Zsb counterterm where the photon vertex goes with Qd, and the
% value of Qd can be considered arbitrary here (so long as Qu = 1 - Qd). Thus,
%
% X_gamma = (B-C) Qd
%
% For the Z bosons, we get (apart from the overall e -> gZ = e/(cw sw) replacement):
%
% X_Z = A ( 1/2 - Qu sw^2) + B ( -1/2 - Qd sw^2 ) + C cw^2 = ( B - C ) ( -1/2 - Qd sw^2 ),
%
% where we have used Qu = 1 - Qd, A + C = 0, and sw^2 + cw^2 = 1 to get r.h.s.
%
% Thus, the ratio X_Z/X_gamma ensures that when X_gamma gets renormalized away by the
% Zsb counterterm, then X_Z gets automatically renormalized away, too. 

\subsection{Results}

With our calculation, we have confirmed the one- and two-loop results 
from Ref.~\cite{Buchalla:1992zm} which are given by (for $\epsilon \to 0$)
\begin{eqnarray}
  C_A^{Z,t,(0)}(\mu_0) &=& \frac{-6x+x^2}{16(x-1)} + \frac{2x+3x^2}{16 (x-1)^2} \ln x
  \,,\nonumber\\[2mm]
  C_A^{Z,t,(1)}(\mu_0) &=& 
     \frac{29x+7x^2+4x^3}{6(x-1)^2}
   - \frac{23x+14x^2+3x^3}{6(x-1)^3} \ln x 
   - \frac{4x+x^3}{2(x-1)^2} \,\mbox{Li}_2\left(1-\frac{1}{x}\right)\nonumber \\[1mm]
&& +\ln\left(\frac{\mu_0^2}{m_t^2}\right) \left[  \frac{8x+x^2+x^3}{2(x-1)^2}
   - \frac{x+4x^2}{(x-1)^3}  \ln x \right]\,.
\end{eqnarray}
Furthermore, similarly to $C_A^W$, we
obtain exact dependence on $M_W$ and $m_t$ for the $\mu_0$-dependent terms 
which read
\begin{eqnarray}
  C_A^{Z,t,(2)}(\mu_0) &=& 
  C_A^{Z,t,(2)}(\mu_0 = m_t) 
 + \ln\left(\frac{\mu_0^2}{m_t^2}\right) \left[
\frac{188x+4x^2+95x^3-47x^4}{6 (x-1)^3} \,\mbox{Li}_2\left(1-\frac{1}{x}\right)\right. \nonumber \\[2mm]
  &+&\!\! \frac{1468x+1578x^2-25x^3-141x^4}{18(x-1)^4} \ln x
\left. - \frac{4622x+1031x^2+582x^3-475x^4}{36(x-1)^3} \right] \nonumber \\[2mm] 
  &+&\!\! \ln^2\left(\frac{\mu_0^2}{m_t^2}\right) \left[ \frac{49x+315x^2-4x^3}{6(x-1)^4} \ln x 
- \frac{440x+257x^2+72x^3-49x^4}{12(x-1)^3} \right].
\end{eqnarray}
For the generic three-loop contributions,
terms up to order $y^{12}$ and $w^{16}$ have been evaluated, as in the 
$W$-box case in Section~\ref{sec::WBox}.
In a numerical form, they read
\begin{eqnarray}
  C_A^{Z,t,(2)}(\mu_0 = m_t) &=& 
\frac{0.1897}{y^2}
+ 2.139
+ 28.59 \, y^2
+ 33.85 \, y^2 \ln y
+ 28.01 \, y^4
+ 97.98 \, y^4 \ln y \nonumber \\
&& - 31.41 \, y^6
+ 106.2 \, y^6 \ln y
- 167.0 \, y^8
- 78.59 \, y^8 \ln y
- 387.4 \, y^{10} \nonumber \\
&& - 618.3 \, y^{10} \ln y
- 697.9 \, y^{12}
- 1688. \, y^{12} \ln y
+ {\mathcal O}\left(y^{14}\right)
  \,, \nonumber\\
  C_A^{Z,t,(2)}(\mu_0 = m_t) &=&
- 1.934
+ 0.8966 \, w
+ 0.7399 \, w^2
+ 0.6058 \, w^3
+ 0.5113 \, w^4
+ 0.4439 \, w^5 \nonumber \\
&& + 0.3948 \, w^6
+ 0.3582 \, w^7
+ 0.3303 \, w^8
+ 0.3087 \, w^9
+ 0.2916 \, w^{10} \nonumber \\
&& + 0.2778 \, w^{11}
+ 0.2667 \, w^{12}
+ 0.2575 \, w^{13}
+ 0.2498 \, w^{14}
+ 0.2433 \, w^{15} \nonumber \\
&& + 0.2379 \, w^{16} 
+ {\mathcal O}\left(w^{17}\right)
  \,.
  \label{eq::CAZt}
\end{eqnarray}

For the fermion triangle contributions, all the $\ln \mu_0$
contributions have cancelled out after matching. We find
\begin{eqnarray}
  C_A^{Z,t,\rm tria.} &=& 
-\frac{0.9871}{y^2}
- 2.388
-1.627 \, y^2
- 3.516 \, y^2 \ln y
-1.830 \, y^4
-6.959 \, y^4 \ln y\nonumber \\
&& -2.038 \, y^6
-10.83 \, y^6 \ln y
-2.210 \, y^8
-15.09 \, y^8 \ln y
-2.353 \, y^{10}
-19.65 \, y^{10} \ln y\nonumber \\
&& -2.473 \, y^{12}
-24.48 \, y^{12} \ln y + {\mathcal O}\left(y^{14}\right)\,, \nonumber \\
  C_A^{Z,t,\rm tria.} &=& -2.418
-1.334\, w
-1.147\, w^2
-1.080\, w^3
-1.048\, w^4
-1.030\, w^5
-1.019\, w^6\nonumber \\
&& -1.012\, w^7
-1.007\, w^8
-1.003\, w^9
-1.001\, w^{10}
-0.9984\, w^{11}
-0.9968\, w^{12}\nonumber \\
&& -0.9955\, w^{13}
-0.9944\, w^{14}
-0.9936\, w^{15}
-0.9928\, w^{16}
+ {\mathcal O}\left(w^{17}\right)\,, \nonumber \\
  C_A^{Z,c,\rm tria.} &=& -1.250
+1.500 \ln y
-0.5331 \, y^2
+0.2778 \, y^2 \ln y
-0.2222 \, y^2 \ln^2 y
+0.1144 \, y^4\nonumber \\
&& -0.08194 \, y^4 \ln y
+0.02778 \, y^4 \ln^2 y
-0.003538 \, y^6
+0.002143 \, y^6 \ln y\nonumber \\
&& -0.0001573 \, y^8
+0.0001235 \, y^8 \ln y
-0.00001283 \, y^{10}
+0.00001145 \, y^{10} \ln y\nonumber \\
&& -1.383 \cdot 10^{-6} \, y^{12}
+ 1.338 \cdot 10^{-6} \, y^{12} \ln y
+ {\mathcal O}\left(y^{14}\right)\,, \nonumber \\
  C_A^{Z,c,\rm tria.} &=&
-1.672
-0.5336 \, w
-0.3100 \, w^2
-0.2181 \, w^3
-0.1683 \, w^4
-0.1370 \, w^5 \nonumber \\
&& -0.1156 \, w^6
-0.09997 \, w^7
-0.08808 \, w^8
-0.07873 \, w^9
-0.07118 \, w^{10}\nonumber \\
&& -0.06495 \, w^{11}
-0.05973 \, w^{12}
-0.05529 \, w^{13}
-0.05147 \, w^{14}
-0.04814 \, w^{15}\nonumber \\
&& -0.04522 \, w^{16}
+ {\mathcal O}\left(w^{17}\right)
  \,.
  \label{eq::CAZtanom}
\end{eqnarray}

\begin{figure}[t]
  \begin{center}
      \includegraphics[width=0.48\textwidth]{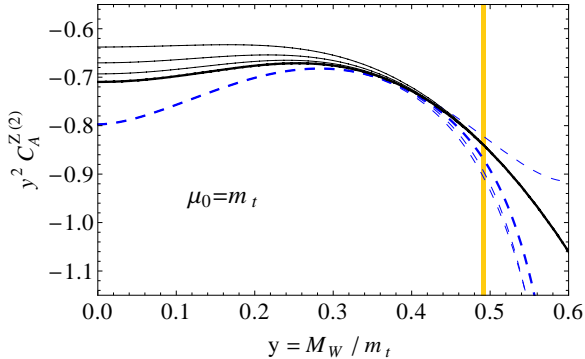}
     \caption{
      \label{fig::result_Z_penguin}
      $y^2 \, C_A^{Z,(2)}$ as a function of $y=M_W/m_t$.
      The (blue) dashed lines are obtained in the 
      limit $y\ll 1$ and the (grey) solid line for $w=1-y^2\ll 1$.
      Thinner lines contain less terms in the expansions. The physical region
      for $y$ is indicated by the (yellow) vertical band.}
  \end{center}
\end{figure}

In the limit of large top quark mass, the coefficient $C_A^{Z,(2)}$
grows as $m_t^2$, which has its origin in the Yukawa interaction
of the charged pseudo-goldstones with the top quark. For this 
reason, we plot in Fig.~\ref{fig::result_Z_penguin} the combination
$y^2 \, C_A^{Z,(2)}$ where sums of the results from
Eqs.~(\ref{eq::CAZt}) and~(\ref{eq::CAZtanom}) are shown as dashed ($M_W\ll
m_t$) and solid lines ($M_W\approx m_t$).  Note that after multiplication 
by $y^2$, the latter is expanded in $w=1-y^2$.  Again,
one observes that the two approximations coincide for $y\approx 0.4$,
which suggests that a combination of the two expansions covers the whole
range between $y=0$ and $y=1$. In the physical region, the expansion around
$M_W= m_t$ provides an excellent approximation.  It is interesting to
note that the fermion triangle contribution is more than an order of
magnitude larger than $C_A^{Z,t,(2)}$. This is particularly true for the
physical value of $y$ where we have $y^2 \, C_A^{Z,t,(2)}\approx -0.02$.

A handy approximation formula which works to better than 1\% for $0.3 < y < 0.7$ 
reads
\begin{equation}
  C_A^{Z,(2)}(\mu_0 = m_t)~ 
  \simeq~ 36.802 \, y^3 - 79.060 \, y^2 + 57.988 \, y - 17.222
  \,.
\end{equation}

%- }}}

%- {{{ Numerical analysis

\section{\label{sec::num}Numerical analysis}

In this section, we shall discuss numerical effects of our three-loop QCD
corrections.  The ${\mathcal B}(B_s\to \mu^+ \mu^-)$ branching ratio in the SM
is proportional to $|C_A|^2$ (cf. Eq.~(\ref{eq::BR}) with
$C_S=C_P=0$ and $F_P=1$). Here, we shall consider $|C_A|^2$ only. Evaluation of the
branching ratio itself is relegated to a parallel article~\cite{Bobeth:2013uxa}
where also the new two-loop EW corrections~\cite{Bobeth:2013tba} are
included.

The relevant parameters are as follows.  For the gauge boson masses, we take
$M_Z = 91.1876\,$GeV~\cite{Beringer:1900zz} and $M_W = 80.358\,$GeV
(calculated from $G_F$, $M_Z$ and $\alpha_{em}$). For the strong coupling,
$\alpha_s(M_Z) = 0.1184$ in the five-flavour QCD is
used~\cite{Beringer:1900zz}. Four-loop renormalization group equations (RGE)
are applied to evolve it to other scales. For the top-quark mass, our input is
$M_t = 173.1\,$GeV~\cite{Beringer:1900zz} which we treat as the pole
mass. We convert it to the $\overline{\rm MS}$ scheme with respect to QCD, but
include no shift due to the EW interactions.  This means that our $m_t$ should
be understood as renormalized on-shell with respect to the EW interactions. As
far as QCD is concerned, we use a three-loop relation for converting $M_t$ to
$m_t(m_t)$, which gives $m_t(m_t) \simeq 163.5\,$GeV.  Next, four-loop RGE are
used to find $m_t(\mu_0)$ at other values of $\mu_0$.
\begin{figure}[t]
  \begin{center}
      \includegraphics[width=0.48\textwidth]{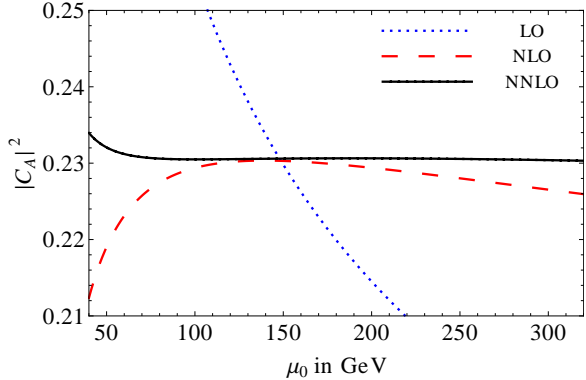}
     \caption{
      \label{fig::cAsq}
      Matching scale dependence of $\left|C_A \right|^2$ and the LO, NLO and NNLO in QCD but 
      at the LO in EW interactions. The top quark mass is renormalized on shell
      with respect to the EW interactions, and at $\mu_0$ in $\overline{\rm MS}$ with respect to QCD.}
  \end{center}
\end{figure}

Fig.~\ref{fig::cAsq} shows the matching scale dependence of 
$\left|C_A \right|^2$. The dotted, dashed and solid curves show the leading order (LO),
next-to-leading-order (NLO) and next-to-next-to-leading-order (NNLO) results,
respectively. In the current case, they correspond to one-, two- and
three-loop matching calculations. 

The range of the plot corresponds roughly to $\mu_0 \in \left[\frac12 M_W, 2
m_t\right]$, which might be considered reasonable given that both the
$W$-boson and the top quark are decoupled simultaneously. However, our Wilson
coefficient has a trivial RGE (at the LO in EW interactions) but it is quite
sensitive to $m_t$. In consequence, the main reason for its
$\mu_0$-dependence here is the top-quark mass renormalization. Thus, for
estimating uncertainties due to truncation of the QCD perturbation series 
at each order, we shall use a more narrow range 
$\mu_0 \in \left[\frac12 m_t, 2 m_t\right]$.

One observes in Fig.~\ref{fig::cAsq} that the prediction for $|C_A|^2$ has
already improved a lot after including the NLO QCD corrections.  The variation
at the NLO level amounts to around $1.8\%$ only, for $\mu_0 \in \left[\frac12
m_t, 2 m_t\right]$. Once the new three-loop corrections are taken into
account, the uncertainty gets reduced to less than $0.2\%$, which can be
treated as negligible for all practical purposes.

Another conclusion that can be drawn from Fig.~\ref{fig::cAsq} is that for~
$\mu_0=m_t$~ the NLO correction is moderate ($2.2\%$), while the NNLO
correction essentially vanishes. Although $\mu_0=m_t$ has been
anticipated to be an optimal scale in the past~\cite{Buchalla:1993bv}, there
has been no convincing theoretical argument for such a choice. Our explicit
three-loop calculation has been actually necessary to suppress the QCD
matching uncertainties in $\left|C_A \right|^2$ to the current sub-percent
level.

For $\mu_0=160\,$GeV, our final result for $C_A$ is well
approximated by the fit
\begin{equation} \label{eq::final.fit}
C_A  = 0.4802 \left( \frac{M_t}{173.1} \right)^{1.52}
                             \left( \frac{\alpha_s(M_Z)}{0.1184} \right)^{-0.09}
                      + {\mathcal O}(\alpha_{em})\,,
\end{equation}
which is accurate to better than $0.1\%$ for 
$\alpha_s(M_Z) \in [0.11,\,0.13]$ and
$M_t \in [170,\,175]\,$GeV.

Let us stress that the ${\mathcal O}(\alpha_{em})$ term in
Eq.~(\ref{eq::final.fit}) stands for both the NLO EW matching corrections at
$\mu = \mu_0$, as well as effects of the evolution of $C_A$ down to $\mu =
\mu_b \sim m_b$, according to the RGE. Once the QCD logarithms get resummed,
the latter effects behave not only like ${\mathcal O}(\alpha_{em})$, but also
like ${\mathcal O}(\alpha_{em}/\alpha_s)$ and ${\mathcal
O}(\alpha_{em}/\alpha_s^2)$, which means that they are potentially more
important than the NNLO QCD corrections evaluated here. However, the actual
numerical impact of the ${\mathcal O}(\alpha_{em}/\alpha_s)$ and ${\mathcal
O}(\alpha_{em}/\alpha_s^2)$ terms on the decay rate amounts to around $-1.5\%$
only~\cite{Misiak:2011bf}, which has been checked using anomalous dimension
matrices from Refs.~\cite{Bobeth:2003at,Huber:2005ig}. The necessary expressions 
are given in the Appendix.

As far as the NLO EW matching corrections are concerned, they have been known
for a long time only in the $m_t \gg M_W$ limit~\cite{Buchalla:1997kz}. A
complete calculation of these two-loop corrections has recently been
finalized~\cite{Bobeth:2013tba}. Their numerical effect depends on the scheme used at
the LO. A detailed discussion of this issue is presented in
Ref.~\cite{Bobeth:2013tba}. Let us only mention that the semi-perfect stabilization of
$\mu_0$-dependence in Fig.~\ref{fig::cAsq} at the NNLO in QCD takes place only
because we have renormalized $m_t$ and $M_W$ on shell with respect to the EW
interactions. If we used $\overline{\rm MS}$ at $\mu_0$ for the EW renormalization of $m_t$
and $M_W$, then acceptable stability would be observed only after including
the very two-loop EW corrections.

%- }}}

%- {{{ Conclusions:

\section{Conclusions}\label{sec::conclusions}

We have evaluated the NNLO QCD corrections to the Wilson coefficient $C_A$
that param-etrizes the $B_s \to \mu^+ \mu^-$ branching ratio in the SM. For this
purpose, three-loop matching between the SM and the relevant effective theory
has been performed. Tadpole integrals depending on $m_t$ and $M_W$ have been
evaluated with the help of expansions starting from the limits $m_t \approx
M_W$ and $m_t \gg M_W$, which for all practical purposes is equivalent to an
exact calculation. When masses of the light quarks and their momenta are set
to zero, care has to be taken about the so-called evanescent operators,
similarly to the NLO case~\cite{Misiak:1999yg}. Such operators have also been
helpful in dealing with diagrams where $\gamma_5$ was present under traces.

Our results for the renormalized matching coefficients $C_A^W$ and
$C_A^Z$ can be downloaded in a computer-readable form
from~\cite{progdata}.  Including the new corrections makes $C_A$ more
stable with respect to the matching scale $\mu_0$ at which the top-quark mass
and $\alpha_s$ are renormalized. Apart from $B_s \to \mu^+ \mu^-$, our
calculation is directly applicable to all the $B_{s(d)} \to \ell^+ \ell^-$
decay modes, and it matters for other processes mediated by $Z$-penguins and
$W$-boxes, e.g., $\bar B \to X_s \nu\bar\nu$, $K \to \pi \nu\bar\nu$, or
short-distance contributions to $K_L \to \mu^+\mu^-$. However, it is only $B_s
\to \mu^+ \mu^-$ for which the three-loop accuracy is relevant at present.

An updated SM prediction for ${\mathcal B}(B_s \to \mu^+ \mu^-)$ is
presented in a parallel article~\cite{Bobeth:2013uxa} where also the new two-loop EW
corrections~\cite{Bobeth:2013tba} are included.

%- }}}

\section*{Acknowledgements}

We thank Alexander Kurz for providing to us his {\tt FORM} routine to compute
tensor tadpole integrals to three loops. We are grateful to
Christoph~Bobeth, Martin~Gorbahn and Emmanuel~Stamou for helpful discussions
and comments on the manuscript. This work was supported by the DFG through
the SFB/TR~9 ``Computational Particle Physics'' and the Graduiertenkolleg
``Elementarteilchenphysik bei h\"ochster Energie und h\"ochster
Pr\"azision''. M.M.  acknowledges partial support by the National Science
Centre (Poland) research project, decision DEC-2011/01/B/ST2/00438.

%- {{{ Appendix:

\section*{Appendix: Logarithmically enhanced QED corrections}

In this appendix, we present explicit expressions for the
logarithmically-enhanced QED corrections to $C_A$. Beyond the LO in
$\alpha_{em}$, its perturbative expansion at the matching scale $\mu_0$
reads 
\begin{equation} 
  C_A(\mu_0) \;=\; C_A^s \;+\; \frac{\alpha_{em}(\mu_0)}{4\pi}\, C_A^{e,(1)}(\mu_0) 
  \;+\; {\mathcal O}(\alpha_{em}^2, \alpha_{em}\alpha_s)\,, 
\end{equation}
where $C_A^s$ stands for the scale-independent ${\mathcal O}(\alpha_{em}^0)$
contribution as given in Eq.~(\ref{eq::cAexp}). Using the RGE from
Refs.~\cite{Bobeth:2003at,Huber:2005ig} one obtains the following result at the
scale $\mu_b$
\begin{eqnarray} \label{eq::cAmub}
  C_A(\mu_b) &=& C_A^s \;+\; 
  \frac{\alpha_{em}(\mu_b)}{\alpha_s^2(\mu_b)}\, F_1 \,\sin^2 \theta_W \;+\;
  \frac{\alpha_{em}(\mu_b)}{\alpha_s  (\mu_b)}\,\left[ F_2 \,+\, F_3 \, \sin^2 \theta_W \right]\nonumber\\[2mm]
  && +\; \alpha_{em}\, G \;+\; {\mathcal O}\left(\frac{\alpha_{em}^2}{\alpha_s^3},\alpha_{em}\alpha_s \right), 
\end{eqnarray}
where $G$ includes all the NLO EW corrections that are not logarithmically enhanced.
The quantities $F_{1,2,3}$ depend on $\eta = \frac{\alpha_s(\mu_0)}{\alpha_s(\mu_b)}$ and 
$x = \frac{m_t^2}{M_W^2}$. We find
\begin{eqnarray} 
F_1 &=& \sum_{i=1}^8 p_i \eta^{a_i}\,,\nonumber\\[2mm]
F_2 &=& \frac{3(\eta-1)}{23\,\eta}\, Y(x)\,,\nonumber\\[2mm]
F_3 &=& \frac{3(\eta-1)}{23\,\eta}\, V(x) + \frac{z \ln\eta}{\eta} + 
\sum_{i=1}^8 \eta^{a_i} \left[ q_i + \eta\, r_i + \eta\, E(x) s_i 
+ \eta\, t_i \ln\left(\frac{\mu_0^2}{M_W^2}\right) \right]\,,\label{eq::Fi}
\end{eqnarray}
with $z \simeq 0.0553$ and the remaining coefficients summarized in Table~\ref{tab::magnum}.
The functions $Y(x)$, $V(x)$ and $E(x)$ originate from the one-loop SM matching
conditions~\cite{Inami:1980fz}
for various operators in the effective theory. They read
%
% L = \ln \mu_0^2/M_W^2
% C_2^{(0)} = 1
% C_1^{(1)} = 15 + 6 L
% C_4^{(1)} = E(x)  - 2/3 + 2/3 L
% C_9^{(0)} = 1/s_w^2 Y(x) + V(x) + 4/9 - 4/9 L
%             => V(x) = -D(x) - 4 C(x)
%
\begin{eqnarray} 
Y(x) &=& \frac{3 x^2 }{ 8 (x-1)^2} \ln x ~+~ \frac{ x^2 - 4 x }{ 8 (x-1)}\,,\nonumber\\[2mm]
V(x) &=& \frac{ -24 x^4 +  6 x^3 +  63 x^2 -  50 x + 8 }{ 18 (x-1)^4 } \ln x ~+~
      \frac{ -18 x^4 + 163 x^3 - 259 x^2 + 108 x }{ 36 (x-1)^3 }\,,\nonumber\\[2mm]
E(x) &=& \frac{ -9 x^2 + 16 x - 4 }{6 (1-x)^4} \ln x ~+~ \frac{ x^3 + 11x^2 - 18 x }{12 (x-1)^3}\,.
\end{eqnarray}
\begin{table}[t]
\newlength{\minus}
\settowidth{\minus}{$-$}
\newcommand{\m}{\hspace{\minus}}
\newcommand{\n}{\hspace{8mm}}
\begin{center}
$\begin{array}{|c|llllllll|}\hline
i   &\n   1   &\n   2   &\n    3       &\n    4           &\n 5 &\n 6 &\n 7  &\n 8 \\\hline
&&&&&&&&\\[-4mm]
a_i &  -2      &  -1      &\n\frac{6}{23}&\m -\frac{12}{23} & 0.4086 &  -0.4230 &-0.8994 &\m 0.1456\\[1mm]
p_i &  -0.0222 &  -0.0768 &   -0.0714    &\m  0.0672        & 0.0074 &\m 0.0360 &\m 0.0614 &  -0.0014\\
q_i &\m 0      &\m 0      &\m  0.2440    &   -0.2231        & 0.1204 &  -0.2874 &  -0.3080 &  -0.0429\\
r_i &\m 0.4464 &\m 0.1626 &   -0.0116    &   -0.0316        & 0.0027 &  -0.0299 &  -0.0421 &\m 0.0004\\
s_i &\m 0.0040 &\m 0.0183 &\m  0         &\m  0             & 0.0017 &\m 0.0076 &  -0.0320 &\m 0.0004\\
t_i &\m 0.0271 &\m 0.0469 &   -0.0114    &   -0.0214        & 0.0018 &  -0.0093 &  -0.0337 &  -0.0001\\\hline
\end{array}$
\caption{\sf Powers and coefficients in Eq.~(\ref{eq::Fi}). \label{tab::magnum}}
\end{center}
\end{table}

The coefficients in Table~\ref{tab::magnum} satisfy the following identities:
\begin{equation} 
\sum_{i=1}^8 p_i a_i ~=~ \sum_{i=1}^8 p_i ~=~ \sum_{i=1}^8 (q_i + r_i) ~=~
\sum_{i=1}^8 s_i ~=~ \sum_{i=1}^8 t_i ~=~ 0.
\end{equation}
With the help of them one can easily check that the terms in
Eq.~(\ref{eq::cAmub}) proportional to $F_{1,2,3}$ are finite in the
limit $\alpha_s \to 0$. Logarithms $\ln^n \frac{\mu_0^2}{\mu_b^2}$ with
$n=1,2$ arise in this limit, which explains why the corresponding QED
corrections are called logarithmically enhanced. The QED logarithms
$\alpha_{em} \ln \frac{\mu_0^2}{\mu_b^2}$ are not being resummed here. It is
the resummation of QCD logarithms $\alpha_s \ln \frac{\mu_0^2}{\mu_b^2}$ in
the corresponding terms that brings inverse powers of $\alpha_s$ into the
final results.

%- }}}

\end{document}